\documentstyle[eqsecnum,preprint,aps]{revtex}

\begin{document}
\input FEYNMAN
\bigphotons

\draft

\title{Gauge Invariance in Nonlocal Regularized QED}
\author{M. A. Clayton}
\address{Department of Physics\\
University of Toronto\\
Toronto, Ontario,\\
Canada M5S 1A7}

\date{\today}
\maketitle

\begin{abstract}
The existence of the invariant measure in nonlocal regularized actions is
discussed. It is shown that the measure for nonlocally regularized QED, as
presented in\cite{Moff-Wood}, exists to all orders, and is precisely what is
required to maintain gauge invariance at one loop and guarantees perturbative
unitarity.
We also demonstrate how the given procedure breaks down in anomalous theories,
and discuss its generalization to other actions.
\end{abstract}

\hbox{UTPT-93-14}
\narrowtext

\section*{Introduction}

The nonlocal regularization scheme introduced recently\cite{Moff-Wood} relies
heavily on the existence of an invariant path integral measure to ensure that
the
extended nonlocal gauge invariance is respected in the quantum regime, thereby
maintaining perturbative unitarity.
Operationally, the measure is expected to correct a Lagrangian generated loop
graph so that the process in question satisfies the Ward identities.
It is not surprising then, that one can directly relate the measure
contribution
to these identities.
This not only guarantees the existence of the measure (through existence of the
Lagrangian generated graphs), but also produces an obvious minimal choice,
which
is necessary to ensure that the scheme does not produce arbitrary results.

In nonlocal regularization of QED, it has been argued previously that the
invariant measure exists to one loop\cite{Moff-Wood}, but without any explicit
results beyond second order.
Comparisons have also been made between nonlocal regularized amplitudes and the
corresponding dimensionally regulated ones, to infer the form of a possible
measure contribution\cite{KW-YM}.
The result has not been proven in general, and we feel that for a true
understanding of nonlocal regularization, one should not have to resort to
another regularization scheme.
We will prove here the relationship between the measure and Ward identities in
QED, and discuss how the result will also hold in other actions.

In section \ref{reg}, we introduce the nonlocal regulated action and the
measure
consistency conditions.
The Ward identities are developed in Sect. \ref{ward}, followed by second,
third
and fourth order corrections in Sections \ref{vp}, \ref{vert} and \ref{box},
respectively.
Next we discuss gauge invariance of higher loop corrections, and explicitly
demonstrate it for the two loop vacuum polarization corrections in Sect.
\ref{twoloop}.
Finally, in Section \ref{anom} possible barriers to consistent quantization are
discussed.
In an Appendix, we prove that the measure exists to all orders, and is exactly
what is necessary in order to satisfy the Ward identities to one loop.

\section{Regularization}
\label{reg}

\subsection{Nonlocal Action}

The standard gauge invariant Lagrangian for local QED is written as
\begin{eqnarray}\label{loclag}
L&=&{\bar\psi}(i{\hbox{$\partial\mkern-10mu/\mkern 1mu$}}-m)\psi
-\frac{1}{4e^2}F^2-
{\bar\psi}{\hbox{$A\mkern-10mu/\mkern 1mu$}}\psi\nonumber \\
&\equiv&{\bar\psi} S^{-1}\psi+\frac{1}{2e^2}A^\mu
D^{-1}_{\mu\nu}A^\nu -{\bar\psi}{\hbox{$A\mkern-10mu/\mkern 1mu$}}\psi,
\end{eqnarray}
which possesses the infinitesimal invariance:
\begin{equation}\label{gi}
\delta A^\mu=\partial^\mu\theta,\quad\delta\psi=-i\theta\psi,
\end{equation}
where we have introduced the inverse propagators into the Lagrangian in
order to clarify notation later.
Gauge fixing is then implemented via the introduction of:
\begin{equation}
L_{GF}=-\frac{1}{2\xi}(\partial_\mu A^\mu)^2,
\end{equation}
leading to the (now invertible) photon propagator:
\begin{equation}
iD_{\mu\nu}=\frac{i}{\Box}(g_{\mu\nu}-
(1-\xi)\frac{\partial_\mu\partial_\nu}{\Box}).
\end{equation}

In order to produce a regulated action that is gauge invariant, we begin by
introducing two types of smeared fermion propagators and give their Schwinger
parameterized form (useful for explicit calculations but not used extensively
in
the present work):
\begin{eqnarray}\label{props}
\hat{S}(p)&=&E_m^2(p)S(p)=-\int_1^\infty\frac{dx}{\Lambda^2}
exp(x\frac{p^2-m^2}{\Lambda^2})({\hbox{$p\mkern-9mu/\mkern 0mu$}} +m),\nonumber
\\
\bar{S}(p)&=&(1-E_m^2(p))S(p)=-\int_0^1\frac{dx}{\Lambda^2}
exp(x\frac{p^2-m^2}{\Lambda^2})({\hbox{$p\mkern-9mu/\mkern 0mu$}}+m),
\end{eqnarray}
(note that $\hat{S}+\bar{S}=S$, the local propagator) and for the photon:
\begin{equation}
\hat{D}_{\mu\nu}=E_0^2(P^2)D_{\mu\nu}
=\int_1^\infty\frac{dx}{\Lambda^2}exp(x\frac{p^2}{\Lambda^2})
(g_{\mu\nu}-(1-\xi)\frac{p_\mu p_\nu}{p^2}),
\end{equation}
where
\begin{equation}
E_m(p^2)=exp(\frac{p^2-m^2}{2\Lambda^2}).
\end{equation}

We now construct the auxiliary `shadow field' Lagrangian:
\begin{equation}\label{aux1}
L_{Sh}={\bar\psi}\hat{S}^{-1}\psi+{\bar\phi}{\bar S}^{-
1}\phi+\frac{1}{2e^2}A^\mu\hat{D}^{-1}_{\mu\nu}A^\nu-
({\bar\psi}+{\bar\phi}){\hbox{$A\mkern-10mu/\mkern 1mu$}}(\psi +\phi).
\end{equation}
We remind the reader that the shadow fields ($\phi,\bar{\phi}$) are introduced
merely as a device to generate the nonlocal action and symmetries in a compact
form.
They do not represent independent
degrees of freedom, as they are constrained to obey their field equations
at the classical level and are not integrated over in the generating
functional.
This is further demonstrated by the fact that their two-point function (the
`barred' propagator in (\ref{props})) does not have a pole and, hence, they
are not propagating degrees of freedom, and should not be included in
asymptotic states.

As discussed in\cite{me}, this particular choice of Lagrangian corresponds to a
nonlocal regularization of QED, in which the classical theory retains the
smearing on the internal photon lines, and internal fermion lines
are `localized'.
This guarantees decoupling of longitudinal photons from on-shell tree
graphs\cite{Moff-Wood}, however in Section \ref{vert}, we also develop the
action
that corresponds to localizing all the fields at the classical level.
This is not necessary in order to guarantee gauge invariance but it treats all
fields in a symmetrical way, and is another viable regularized QED action.

When performing quantum corrections, convergence is guaranteed by the presence
of the `hatted' propagator on at least one internal line.
We have given the Schwinger parameterized form of the propagators in
(\ref{props}), since in practice one writes the local graph in Schwinger
parameter form and restricts the range of parameter integrals appropriate for
the
process in question\cite{KW-2L}.
(For example, when one calculates single fermion loop graphs, the unit
hypercube is removed from the volume of integration.
This corresponds to the absence of the contribution from the graph with all
`barred' internal lines.)
Clearly (\ref{aux1}) is invariant under (the BRST generalization of):
\begin{eqnarray}\label{nlgi}
\delta A^\mu&=&\partial^\mu\theta\nonumber \\
\delta\psi&=&-iE^2\theta(\psi+\phi),\nonumber \\
\delta\phi&=&-i(1-E^2)\theta(\psi+\phi),
\end{eqnarray}
and the conserved Noether current (generalizing the local vector
current) is given by
\begin{equation}
J^\mu=(\bar{\psi}+\bar{\phi})\gamma^\mu(\psi+\phi).
\end{equation}
To generate the action in terms of physical fields alone, we must
remove them from the classical action by forcing them to obey their
classical equations of motion.
We have:
\begin{equation}
\phi=\bar{S}{\hbox{$A\mkern-10mu/\mkern 1mu$}}(\psi
+\phi)=(1-\bar{S}{\hbox{$A\mkern-10mu/\mkern 1mu$}})^{-1}
\bar{S}{\hbox{$A\mkern-10mu/\mkern 1mu$}}\psi,
\end{equation}
and the Lagrangian, gauge transformations and Noether current are
then given by:
\begin{eqnarray}\label{expand}
L_{NL}&=&\bar{\psi}\hat{S}^{-1}\psi+\frac{1}{2e^2}A^\mu\hat{D}^{-
1}_{\mu\nu}A^\nu-\bar{\psi}{\hbox{$A\mkern-10mu/\mkern
1mu$}}(1-\bar{S}{\hbox{$A\mkern-10mu/\mkern 1mu$}})^{-1}\psi,\nonumber \\
&=&\bar{\psi}\hat{S}^{-1}\psi+\frac{1}{2e^2}A^\mu\hat{D}^{-
1}_{\mu\nu}A^\nu-\bar{\psi}\hbox{$A\mkern-10mu/\mkern 1mu$}\psi,\nonumber \\
&-&\bar{\psi}\hbox{$A\mkern-10mu/\mkern 1mu$}\bar{S}\hbox{$A\mkern-10mu/\mkern
1mu$}\psi
-\bar{\psi}\hbox{$A\mkern-10mu/\mkern 1mu$}\bar{S}\hbox{$A\mkern-10mu/\mkern
1mu$}\bar{S}\hbox{$A\mkern-10mu/\mkern 1mu$}\psi
-\bar{\psi}\hbox{$A\mkern-10mu/\mkern 1mu$}\bar{S}\hbox{$A\mkern-10mu/\mkern
1mu$}\bar{S}\hbox{$A\mkern-10mu/\mkern 1mu$}\bar{S}\hbox{$A\mkern-10mu/\mkern
1mu$}\psi-\ldots,
\end{eqnarray}
\begin{eqnarray}\label{pexp}
\delta\psi&=&-iE^2\theta(1-\bar{S}{\hbox{$A\mkern-10mu/\mkern 1mu$}})^{-1}\psi
\nonumber \\
&=&-iE^2\theta(1+\bar{S}\hbox{$A\mkern-10mu/\mkern 1mu$}+\bar{S}\hbox{$A\mkern-
10mu/\mkern 1mu$}\bar{S}\hbox{$A\mkern-10mu/\mkern 1mu$}+\ldots)\psi,
\end{eqnarray}
\begin{eqnarray}
J^\mu&=&\bar{\psi}(1-{\hbox{$A\mkern-10mu/\mkern
1mu$}}\bar{S})^{-1}\gamma^\mu(1-
\bar{S}{\hbox{$A\mkern-10mu/\mkern 1mu$}})^{-1}\psi\nonumber \\
&=&\bar{\psi}(\gamma^\mu+\hbox{$A\mkern-10mu/\mkern
1mu$}\bar{S}\gamma^\mu+\gamma^\mu\bar{S}\hbox{$A\mkern-10mu/\mkern 1mu$}
+\ldots)\psi,
\end{eqnarray}
These results reproduce the classical theory described in\cite{Moff-Wood} (up
to
a rescaling of the electromagnetic field strength
$A\rightarrow -eA$).

\subsection{Generating the Measure}
\label{measure}

Quantizing the theory described by (\ref{expand}) in
the path integral formalism requires finding an invariant measure that respects
the full nonlocal gauge invariance described by (\ref{pexp}), since
the trivial measure is no longer invariant.
We therefore require a method to generate consistency conditions on an
invariant
measure in order to retain the nonlocal invariance in the quantum regime, and
thereby guarantee decoupling.
The simplest way to do this is to consider how the trivial measure transforms
under the nonlocal regularization gauge transformations, and
require that there is a contribution from the measure that compensates.

Writing the full invariant measure as the product of the
trivial measure and an exponentiated action term:
\begin{equation}
\mu_{inv}[\phi]=d\phi\;exp(iS_{meas}[\phi]),
\end{equation}
we find\cite{me}:
\begin{equation}\label{mcon}
\delta S_{meas}=iTr[\frac{\partial}{\partial\phi}\delta\phi].
\end{equation}
This procedure determines the measure up to arbitrary gauge invariant
terms but, as we shall see, there is a natural minimal choice determined
through
relating the measure to the one loop graph it corrects, resulting in a unique
(if
it exists) longitudinal measure.
We feel that any other invariant terms properly belong in the Lagrangian and
should not be introduced into the measure.
The measure is also constrained to be an entire function of the
4-momentum invariants for the particular process, ensuring that no
additional degrees of freedom become excited in the quantum regime.

We then write the expectation of any operator as:
\begin{eqnarray}\label{top}
<T^*[{\cal O}[A^\mu,\psi,\bar{\psi}]]>=\int d\mu_{inv}
{\cal O}[A^\mu,\psi,\bar{\psi}]exp[i\int d^4xL_{NL}],
\end{eqnarray}
and the perturbative expansion is implemented as usual via the generating
functional:
\begin{eqnarray}\label{genfun}
Z[S_\mu,\bar{\eta},\eta]=\int d\mu_{inv}
exp[i\int d^4x(L_{NL}+S_\mu A^\mu+\bar{\eta}\psi+\bar{\psi}\eta)].
\end{eqnarray}

\section{Ward Identities}
\label{ward}

In order to generate identities on n-point functions, one transforms the fields
as in Eq. (\ref{nlgi}), and sets the infinitesimal
variation of the generating functional to zero:
\begin{eqnarray}\label{cond}
Z_0&=&\frac{\delta}{\delta\omega}Z[S_\mu,\bar{\eta},\eta]\mid_{\omega
=0}
=\int
d\mu_{inv}\;K[A^\mu,S^\mu,\bar{\psi},\psi,\bar{\eta},\eta](x)exp(iS[J])=0,
\end{eqnarray}
where $K$ is given by:
\begin{equation}
K=-\frac{1}{e^2\xi}\frac{\Box\partial_\mu}{E^2_0}A^\mu-\partial_\mu S^\mu
-i\bar{\eta}E^2(1-\bar{S}\hbox{$A\mkern-10mu/\mkern 1mu$})^{-1}\psi
+i\bar{\psi}(1-\hbox{$A\mkern-10mu/\mkern 1mu$}\bar{S})^{-1}E^2\eta.
\end{equation}
Setting all sources to zero gives:
\begin{equation}
Z_0\mid_{J=0}=-\frac{1}{e^2\xi}\frac{\Box\partial_\mu}{E^2_0}<T^*[A^\mu(x)]>=0,
\end{equation}
and one derivative with respect to a photon source gives:
\begin{equation}
\frac{1}{i}\frac{\delta}{\delta S_\alpha(y)}Z_0\mid_{J=0}
=-\frac{1}{e^2\xi}\frac{(\Box\partial_\mu)_x}{E^2_0}<T^*[A^\mu(x)A^\alpha(y)]>
-\partial^\alpha\delta(x-y)=0.
\end{equation}
This relation is seen to hold to lowest order as the delta-function term
cancels the longitudinal term in the bare propagator.
It also provides a relation on the the irreducable corrections to the
photon self energy (after truncating the external legs):
\begin{equation}
p_\mu\Pi^{\mu\alpha}=0.
\end{equation}
Higher derivatives with respect to photon sources then gives similar relations:
\begin{equation}
\prod_j\frac{1}{i}\frac{\delta}{\delta S_{\alpha_j}(y_j)}Z_0\mid_{J=0}
=-\frac{1}{e^2\xi}\frac{(\Box\partial_\mu)_x}{E^2_0}
<T^*[A^\mu(x)\prod_jA^\alpha_j(y_j)]>=0,
\end{equation}
which results in the identity on the n-point photon function:
\begin{equation}
p_\mu N^{\mu\alpha_1\ldots\alpha_{n-1}}=0.
\end{equation}

Furthur identities are derived from taking functional derivatives of $Z_0$
with respect to the sources.
For example, one derivative with respect to each of the fermion and antifermion
sources provides (to lowest order):
\begin{eqnarray}
\frac{1}{i}\frac{\delta}{\delta\eta(z)}
\frac{1}{i}\frac{\delta}{\delta\bar{\eta}(y)}Z_0\mid_{J=0}
&=&-\frac{1}{e^2\xi}\frac{(\Box\partial_\mu)_x}{E^2_0}
<T^*[A^\mu(x)\bar{\psi}(z)\psi(y)]>\nonumber \\
&-&(E^2_m\delta^4(x-y))<T^*[\bar{\psi}(z)\psi(x)]>\nonumber \\
&+&(E^2_m\delta^4(x-z))<T^*[\bar{\psi}(x)\psi(y)]>=0,
\end{eqnarray}
leads to the usual identity on the vertex correction:
\begin{equation}\label{vid}
-i(p^\prime-p)_\mu\Gamma^\mu(p,p^\prime)=-ie(S^{-1}(p)-S^{-1}(p^\prime)),
\end{equation}
where we are referring to the fully corrected functions.
Although the form of identities pertaining to higher point graphs are not
simple
in general, it is clear that they guarantee decoupling of photons from graphs
with on-shell external fermions.

\section{Vacuum Polarization}
\label{vp}

We will begin with a brief review of the results derived elsewhere
\cite{Moff-Wood} on vacuum polarization to one loop.
The nontrivial contributions to the measure come from:
\begin{equation}\label{dlpsi}
\delta\psi=-iE^2\theta\bar{S}{\hbox{$A\mkern-10mu/\mkern 1mu$}}\psi,
\end{equation}
and, as given in the original paper, produce the
necessary contribution to vacuum polarization in order to satisfy
the Ward identity and keep the photon transverse:
\begin{equation}\label{vac}
S_{meas}^{(2)}=-\frac{\Lambda^2}{4\pi^2}\int\frac{dp\;dq}{(2\pi)^4}(2\pi)^4
\delta^4(p+q)M_v(p)A^\mu(p)A_\mu(q),
\end{equation}
\begin{equation}
M_v(p)=\int_0^\frac{1}{2}dt(1-t)exp(t\frac{p^2}{\Lambda^2}-
\frac{1}{1-t}\frac{m^2}{\Lambda^2}).
\end{equation}
However, we will now rewrite this result in a form that makes it more obvious
what is happening.
First consider the one loop contribution to vacuum polarization (Fig.
\ref{fvp}):
\begin{equation}
-i\Pi^{\mu\nu}=-\int\frac{dk}{(2\pi)^4}\sum_{Ev}Tr[S(k)\gamma^\nu
S(p+k)\gamma^\mu].
\end{equation}
where the sum represents all terms coming from (\ref{expand}):
\begin{eqnarray}\label{sumv}
\sum_{Ev}&=&E^2_m(k)E^2_m(p+k)+(1-E^2_m(k))E^2_m(p+k)+
E^2_m(k)(1-E^2_m(p+k))\nonumber \\
&=&E^2_m(p+k)(1-E^2_m(k))+E^2_m(k)\nonumber \\
&=&E^2_m(k)(1-E^2_m(p+k))+E^2_m(p+k).
\end{eqnarray}
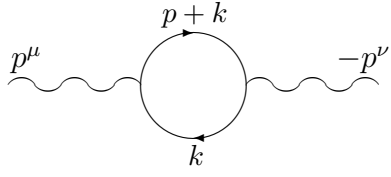
\begin{figure}[p]
\begin{center}
\begin{picture}(10000,10000)
\drawline\photon[\E\REG](0,5000)[5]
\global\advance\photonfronty by 700
\global\advance\photonfrontx by 100
\put(\photonfrontx,\photonfronty){$p^\mu$}
\global\advance\photonbackx by 2000
\put(\photonbackx,\photonbacky){\circle{4000}}
\global\advance\photonbacky by 2000
\drawarrow[\E\ATTIP](\photonbackx,\photonbacky)
\global\advance\photonbacky by 400
\global\advance\photonbackx by -1200
\put(\photonbackx,\photonbacky){\small $p+k$}
\global\advance\photonbacky by -4400
\global\advance\photonbackx by 1200
\drawarrow[\W\ATTIP](\photonbackx,\photonbacky)
\global\advance\photonbacky by -1000
\global\advance\photonbackx by -200
\put(\photonbackx,\photonbacky){\small $k$}
\global\advance\photonbacky by 3000
\global\advance\photonbackx by 2200
\drawline\photon[\E\REG](\photonbackx,\photonbacky)[5]
\global\advance\photonbacky by 700
\global\advance\photonbackx by -1600
\put(\photonbackx,\photonbacky){$-p^\nu$}
\end{picture}
\end{center}
\caption{One loop contribution to vacuum polarization}
\label{fvp}
\end{figure}
We now `dot' $p_\mu$ into this and use the relation (which will recieve heavy
use in this paper);
\begin{equation}
\hbox{$p\mkern-8mu/\mkern 0mu$}=S^{-1}(p+k)-S^{-1}(k).
\end{equation}
Then:
\begin{equation}
-ip_\mu\Pi^{\mu\nu}=-
\int\frac{dk}{(2\pi)^4}\sum_{Ev}(Tr[S(k)\gamma^\nu]-Tr[S(p+k)\gamma^\nu]).
\end{equation}
Re-writing the sums as the last two terms in (\ref{sumv}), one each for
the traces:
\begin{eqnarray}
-ip_\mu\Pi^{\mu\nu}&=&-\int\frac{dk}{(2\pi)^4}
\{E^2_m(p+k)Tr[\bar{S}(k)\gamma^\nu]+Tr[\hat{S}(k)\gamma^\nu]\}\nonumber \\
&+&\int\frac{dk}{(2\pi)^4}
\{E^2_m(k)Tr[\bar{S}(p+k)\gamma^\nu]+Tr[\hat{S}(p+k)\gamma^\nu]\},
\end{eqnarray}
where the second and fourth terms cancel by a simple shift of loop momentum.
Note that we have been careful to only consider partitioning the sum into
separately convergent terms.

We are then left with:
\begin{eqnarray}\label{vprs}
-ip_\mu\Pi^{\mu\nu}&=&-\int\frac{dk}{(2\pi)^4}
\{E^2_m(p+k)Tr[\bar{S}(k)\gamma^\nu]-
E^2_m(k)Tr[\bar{S}(p+k)\gamma^\nu]\}\nonumber \\
&\stackrel{df}{=}&-ip_\mu\Pi^{\mu\nu}_L,
\end{eqnarray}
where:
\begin{equation}
\Pi^{\mu\nu}=(g^{\mu\nu}-\frac{p^\mu p^\nu}{p^2})\Pi_T+
\frac{p^\mu p^\nu}{p^2}\Pi_L,
\quad\Pi^{\mu\nu}_L\stackrel{df}{=}\Pi_Lg^{\mu\nu}.
\end{equation}
{}From the above gauge transformation, we find a condition on the measure that
is
easily identifiable with (\ref{vprs}):
\begin{eqnarray}\label{dsmvp}
\delta S_{meas}^{(2)}&=&-\int\frac{dp\;dq}{(2\pi)^4}(2\pi)^4\delta^4(p+q)
\theta(p)A_\nu(q)\nonumber \\
&&\int\frac{dk}{(2\pi)^4}\{E^2(p+k)Tr[\bar{S}(k)\gamma^\nu]-
E^2(k)Tr[\bar{S}(p+k)\gamma^\nu]\}\nonumber \\
&=&\int\frac{dp\;dq}{(2\pi)^4}(2\pi)^4\delta^4(p+q)
\theta(p)A_\nu(q)(-i)p_\mu\Pi^{\mu\nu}_L(p),
\end{eqnarray}
which leads to:
\begin{equation}\label{vpmeas}
S_{meas}^{(2)}=\frac{1}{2}\int\frac{dp\;dq}{(2\pi)^4}(2\pi)^4\delta^4(p+q)
A_\mu(p)A_\nu(q)\Pi^{\mu\nu}_L(p).
\end{equation}
where we have used the fact that the two point function is symmetrized in the
external fields, even after the longitudinal projection is performed.
A simple calculation will reproduces (\ref{vac}) and we have thus reduced the
existence of the measure to the existence of the longitudinal projection of the
graph.
(i.e. The measure is just what is required in order for the process in question
to satisfy the Ward identities.)
Note that the measure (\ref{vac}) is an entire function of the finite complex
$p^2$ plane and we are therefore sure that we are not introducing additional
degrees of freedom at the quantum level through the measure.
We could also see this directly from the gauge transformations leading to
(\ref{dsmvp}) since the barred propagators do not have a pole, so that when the
resulting measure term is analytically continued to Minkowski space, we will
not
pick up any imaginary parts.

\section{Third Order Corrections}
\label{vert}

\begin{figure}[p]
\begin{center}
\begin{picture}(10000,15000)
\drawline\fermion[\SE\REG](1000,12000)[3000]
\drawarrow[\LDIR\ATTIP](\pmidx,\pmidy)
\global\advance\fermionfrontx by -400
\global\advance\fermionfronty by -1000
\put(\fermionfrontx,\fermionfronty){$p$}
\drawline\photon[\E\REG](\fermionbackx,\fermionbacky)[7]
\drawarrow[\SE\ATTIP](\pmidx,\pmidy)
\global\advance\pmidy by 500
\global\advance\pmidx by -200
\put(\pmidx,\pmidy){\small $k$}
\drawline\fermion[\SE\REG](\fermionbackx,\fermionbacky)[5000]
\drawarrow[\LDIR\ATTIP](\pmidx,\pmidy)
\global\advance\pmidx by -3000
\global\advance\pmidy by -300
\put(\pmidx,\pmidy){\small $p+k$}
\drawline\photon[\S\REG](\fermionbackx,\fermionbacky)[6]
\global\advance\photonbackx by -1500
\global\advance\photonbacky by 300
\put(\photonbackx,\photonbacky){$q^\mu$}
\drawline\fermion[\NE\REG](\fermionbackx,\fermionbacky)[4000]
\drawarrow[\LDIR\ATTIP](\pmidx,\pmidy)
\global\advance\pmidx by 700
\global\advance\pmidy by -300
\put(\pmidx,\pmidy){\small $p^\prime+k$}
\drawline\fermion[\NE\REG](\fermionbackx,\fermionbacky)[4000]
\drawarrow[\LDIR\ATTIP](\pmidx,\pmidy)
\global\advance\fermionbackx by -400
\global\advance\fermionbacky by -1000
\put(\fermionbackx,\fermionbacky){$p^\prime$}
\end{picture}
\end{center}
\caption{Vertex correction}
\label{fvert}
\end{figure}
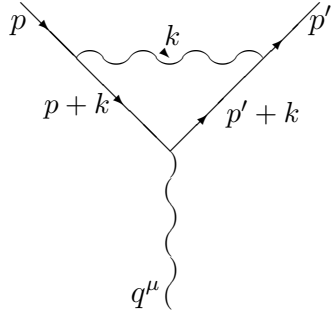

We now check the Ward identity (\ref{vid}) on the vertex correction shown in
Fig.
\ref{fvert}:
\begin{eqnarray}
-i\Gamma^{\mu}(p,p^\prime)&=&
\int\frac{dk}{(2\pi)^4}\sum_{E\Gamma}\gamma^\alpha S(p^\prime+k)
\gamma^\mu S(p+k)\gamma^\beta e^2D_{\alpha\beta}(k),
\end{eqnarray}
where the sum is over all terms in the Lagrangian (\ref{expand}):
\begin{eqnarray}\label{sum}
\sum_{E\Gamma}&=&E^2_0(k)[E^2_m(p+k)E^2_m(p^\prime+k)
+E^2_m(p+k)(1-E^2_m(p^\prime+k))\nonumber \\
&+&(1-E^2_m(p+k))E^2_m(p^\prime+k)
+(1-E^2_m(p+k))(1-E^2_m(p^\prime+k))]=E^2_0(k),
\end{eqnarray}
so that the fermion line is fully localized.
(In this regularization, it is easy to see that in any process througoing
fermion
lines are always fully localized.)

Dotting the photon momentum into this gives:
\begin{eqnarray}
-i(p^\prime-p)_\mu\Gamma^{\mu}(p,p^\prime)&=&
\int\frac{dk}{(2\pi)^4}\gamma^\alpha S(p^\prime+k)
(\hbox{$p\mkern-8mu/\mkern 0mu$}^\prime-\hbox{$p\mkern-8mu/\mkern
0mu$})S(p+k)\gamma^\beta e^2\hat{D}_{\alpha\beta}(k),
\end{eqnarray}
and using the identity: $\hbox{$p\mkern-8mu/\mkern 0mu$}^\prime-\hbox{$p\mkern-
8mu/\mkern 0mu$}=S^{-1}(p^\prime+k)-S^{-1}(p+k)$,
we immediately recognize:
\begin{eqnarray}
-i(p^\prime-p)_\mu\Gamma^{\mu}(p,p^\prime)&=&
\int\frac{dk}{(2\pi)^4}[\gamma^\alpha S(p+k)\gamma^\beta
-\gamma^\alpha S(p^\prime+k)\gamma^\beta]e^2\hat{D}_{\alpha\beta}(k)
\nonumber \\
&=&-i[\Sigma(p)-\Sigma(p^\prime)],
\end{eqnarray}
where
\begin{eqnarray}\label{sig}
-i\Sigma(p)=\int{dk}{(2\pi)^4}\sum_{E\Sigma}
\gamma^\alpha S(p+k)\gamma^\beta e^2D_{\alpha\beta}(k),
\end{eqnarray}
with now
\begin{equation}
\sum_{E\Sigma}=E^2_0(k)[E^2_m(p+k)+(1-E^2_m(p+k))]=E^2_0(k).
\end{equation}
We therefore have to third order:
\begin{equation}
-i(p^\prime-p)\Gamma(p,p^\prime)=
ie[S^{-1}(p)-\Sigma(p)]-ie[S^{-1}(p^\prime)-\Sigma(p^\prime)],
\end{equation}
and the identity (\ref{vid}) is satisfied.

That there is no measure contribution in odd orders was already demonstrated
in \cite{Moff-Wood}, and here we sketch the result for this particular case.
The relavent transformations are:
\begin{eqnarray}
\delta\psi&=&-iE^2_m\theta\bar{S}\hbox{$A\mkern-10mu/\mkern
1mu$}\bar{S}\hbox{$A\mkern-10mu/\mkern 1mu$}\psi,\quad
\delta\bar{\psi}=\bar{\psi}\hbox{$A\mkern-10mu/\mkern
1mu$}\bar{S}\hbox{$A\mkern-
10mu/\mkern 1mu$}\bar{S}\theta E^2_mi,
\end{eqnarray}
leading to the condition on the measure:
\begin{eqnarray}
\delta S^{(3)}_{meas}&=&-\int\frac{dp\;dq_1\;dq_2}{(2\pi)^8}
(2\pi)^4\delta^4(p+q_1+q_2)\int\frac{dk}{(2\pi)^4}\theta(p)\nonumber\\
&\times&\{E^2_m(k+p)Tr[\bar{S}(k)\hbox{$A\mkern-10mu/\mkern 1mu$}(q_1)
\bar{S}(k-q_1)\hbox{$A\mkern-10mu/\mkern 1mu$}(q_2)]\nonumber \\
&-&E^2_m(k)Tr[\hbox{$A\mkern-10mu/\mkern
1mu$}(q_1)\bar{S}(k-q_1)\hbox{$A\mkern-
10mu/\mkern 1mu$}(q_2)\bar{S}(k+p)]\},
\end{eqnarray}
and it is not hard to see that the surviving terms from the traces are of
opposite sign and therefore cancel.
This is consistent with the fact that the related triangle graph disappears by
Furry's theorem, and hence the measure can be related to a projection of which
is zero.

Instead of the regulated Lagrangian (\ref{aux1}) in which only the smearing
of the internal fermion lines is removed at the classical level, we could
also remove the smearing from the internal photon lines.
This is accomplished by introducing a shadow field for the photon as well
as the fermions:
\begin{eqnarray}\label{Bsh}
L_{Sh}&=&{\bar\psi}\hat{S}^{-1}\psi+{\bar\phi}{\bar S}^{-1}\phi
+\frac{1}{2e^2}A^\mu\hat{D}^{-1}_{\mu\nu}A^\nu
+\frac{1}{2e^2}B^\mu\bar{D}^{-1}_{\mu\nu}B^\nu\nonumber \\
&-&({\bar\psi}+{\bar\phi}){\hbox{$A\mkern-10mu/\mkern 1mu$}}(\psi +\phi).
\end{eqnarray}
Then the field equations that are used to remove the shadow fields are:
\begin{eqnarray}\label{eqm}
\phi&=&\bar{S}(\hbox{$A\mkern-10mu/\mkern 1mu$}+\hbox{$B\mkern-12mu/\mkern
2mu$})(\psi+\phi),\nonumber \\
B_\mu&=&e^2\bar{D}_{\mu\nu}(\bar{\psi}+\bar{\phi})\gamma^\nu(\psi+\phi).
\end{eqnarray}
Although we have not been able to give the full nonlocal Lagrangian in closed
form in terms of the physical fields alone, it should be clear that one can
generate it to any order by iterating (\ref{eqm}) in (\ref{Bsh}).

The unitary gauge Lagrangian posesses the gauge invariance:
\begin{eqnarray}\label{blah}
\delta\psi&=&-iE^2_m\theta(\psi+\phi),\quad
\delta\phi=-i(1-E^2_m)\theta(\psi+\phi),\nonumber \\
\delta A^\mu&=&E^2_0\partial^\mu\theta,\quad
\delta B^\mu=(1-E^2_0)\partial^\mu\theta,
\end{eqnarray}
as guaranteed by the nonlocal construction \cite{KW-YM}.
It also has the `dynamically trivial' invariance:
\begin{equation}
\delta A^\mu=(1-E^2_0)\partial^\mu\theta,\quad
\delta B^\mu=-(1-E^2_0)\partial^\mu\theta,
\end{equation}
which allows one to instead write in (\ref{blah}):
\begin{equation}
\delta A^\mu=\partial^\mu\theta,\quad\delta B^\mu=0,
\end{equation}
so that it is clear that longitudinal photons should still decouple.
Indeed, it is not hard to check that the Ward identity is unchanged in this
case,
even though the range of parameter integrals is now different in the
calculation of the vertex correction.
We shall see, however, that there is now a contribution from the measure
that `corrects' the vertex further, allowing consistency with the Ward
identity.

Repeating the above calculation of the vertex correction merely involves
additional terms in the sum (\ref{sum}), which we will now write as:
\begin{equation}
\sum_{E\Gamma}^\prime=E^2_m(k)(1-E^2_m(p+k))(1-E^2_m(p^\prime+k))
+\sum_{E\Sigma}^\prime
\end{equation}
where the second term refers to the sum corresponding to the region
in the fermion correction:
\begin{equation}
\sum_{E\Sigma}^\prime=\sum_{E\Sigma}+(1-E^2_0(k))E^2_m(p+k),
\end{equation}
(we denote all quantities calculated in this `extended' regularization of QED
by primes).
Calculating the vertex correction gives:
\begin{eqnarray}
-i\Gamma^{\mu\prime}&=&\int\frac{dk}{(2\pi)^4}\sum_{E\Gamma}^\prime
e^2\bar{D}_{\alpha\beta}(k)
\gamma^\alpha S(p^\prime+k)\gamma^\mu S(p+k)\gamma^\beta\nonumber \\
-ip_\mu\Gamma^{\mu\prime}&=&\int\frac{dk}{(2\pi)^4},
\{\sum_{E\Sigma}^\prime\gamma^\alpha S(p+k)\gamma^\beta D_{\alpha\beta}(k)-
\sum_{E\Sigma}^\prime\gamma^\alpha S(p^\prime+k)\gamma^\beta
D_{\alpha\beta}(k)\}\nonumber \\
&+&\int\frac{dk}{(2\pi)^4}\bar{D}_{\alpha\beta}(k)
\{E^2_m(p^\prime+k)\gamma^\alpha\bar{S}(p+k)\gamma^\beta
-E^2_m(p+k)\gamma^\alpha\bar{S}(p^\prime+k)\gamma^\beta\}\nonumber \\
&=&-i[\Sigma(p)-\Sigma(p^\prime)]\nonumber \\
&+&\int\frac{dk}{(2\pi)^4}\bar{D}_{\alpha\beta}(k)
\{E^2_m(p^\prime+k)\gamma^\alpha\bar{S}(p+k)\gamma^\beta
-E^2_m(p+k)\gamma^\alpha\bar{S}(p^\prime+k)\gamma^\beta\}\nonumber \\
&\stackrel{df}{=}&-iq_\mu\Gamma^{\mu\prime}_L.
\end{eqnarray}

We determine the measure by first iterating the shadow field equations into the
transformations to find the additional terms at third order;
\begin{eqnarray}
\delta\psi&=&-iE^2_m\theta\bar{S}\gamma^\mu\psi
e^2\bar{D}_{\mu\nu}\bar{\psi}\gamma^\nu\psi,\nonumber \\
\delta\bar{\psi}&=&\bar{\psi}\gamma^\nu\psi
e^2\bar{D}_{\mu\nu}\bar{\psi}\gamma^\mu\bar{S}\theta E^2_mi.
\end{eqnarray}
Note that only the `1PI derivative' need be taken, since the other terms will
merely reproduce lower order measure contributions attached to barred
propagators, serving to localize corrected tree graphs.
This merely implies replacing (\ref{vpmeas}) with:
\begin{eqnarray}
S_{meas}^{(2)\prime}=\frac{1}{2}\int\frac{dp\;dq}{(2\pi)^4}(2\pi)^4\delta^4(p+q)
(A_\mu(p)+B_\mu(p))(A_\nu(q)+B_\nu(q))\Pi^{\mu\nu}_L.
\end{eqnarray}

The remaining contributions result in:
\begin{eqnarray}
\delta S_{meas}^{(3)\prime}&=&\int\frac{dq\;dp\;dp^\prime}{(2\pi)^6}
(2\pi)^4\delta^4(q+p-p^\prime)\frac{dk}{(2\pi)^4}
\theta(q)\bar{D}_{\mu\nu}(k)\nonumber \\
&\times&\{E^2_m(k+p^\prime)\bar{\psi}(-p^\prime)\gamma^\mu
\bar{S}(k+p)\gamma^\nu\psi(p)-E^2_m(k+p)
\bar{\psi}(-p^\prime)\gamma^\mu\bar{S}(k+p^\prime)\gamma^\nu\psi(p)\}
\nonumber \\
&=&\int\frac{dq\;dp\;dp^\prime}{(2\pi)^6}
(2\pi)^4\delta^4(q+p-p^\prime)\theta(q)\bar{\psi}(-p^\prime)
\{-iq_\mu\Gamma^{\mu\prime}_L
-i[\Sigma(p^\prime)^\prime-\Sigma(p)^\prime]\}\psi(p).
\end{eqnarray}
Note that we can write:
\begin{equation}
\Sigma(p^\prime)^\prime-\Sigma(p)^\prime=q_\mu\Sigma^{\mu\prime},
\end{equation}
since the $q\rightarrow 0$ limit implies $p=-p^\prime$,
and so we then have the required measure:
\begin{eqnarray}\label{vmeas}
S_{meas}^{(3)\prime}&=&\int\frac{dq\;dp\;dp^\prime}{(2\pi)^6}
(2\pi)^4\delta^4(q+p-p^\prime)
A_\mu(q)\bar{\psi}(-p^\prime)
\{\Gamma^{\mu\prime}_L+\Sigma^{\mu\prime}\}\psi(p),
\end{eqnarray}
and we see that the measure again ensures the validity of the Ward identity.

\section{Box Graph}
\label{box}

\begin{figure}[p]
\begin{center}
\begin{picture}(10000,15000)
\drawline\photon[\SE\FLIPPED](1000,13000)[6]
\global\advance\pmidy by 100
\global\advance\photonfrontx by -200
\put(\photonfrontx,\pmidy){$p^\mu$}
\drawline\fermion[\E\REG](\photonbackx,\photonbacky)[4000]
\drawarrow[\LDIR\ATTIP](\pmidx,\pmidy)
\global\advance\pmidy by 500
\global\advance\pmidx by -1200
\put(\pmidx,\pmidy){\small $k+p$}
\drawline\photon[\NE\REG](\fermionbackx,\fermionbacky)[6]
\global\advance\pmidy by 100
\global\advance\photonbackx by -600
\put(\photonbackx,\pmidy){$q_3^\gamma$}
\drawline\fermion[\S\REG](\fermionbackx,\fermionbacky)[4000]
\drawarrow[\LDIR\ATTIP](\pmidx,\pmidy)
\global\advance\pmidx by 400
\put(\pmidx,\pmidy){\small $k+p+q_3$}
\drawline\photon[\SE\REG](\fermionbackx,\fermionbacky)[6]
\global\advance\pmidy by -200
\global\advance\photonbackx by -800
\put(\photonbackx,\pmidy){$q_2^\beta$}
\drawline\fermion[\W\REG](\fermionbackx,\fermionbacky)[4000]
\drawarrow[\LDIR\ATTIP](\pmidx,\pmidy)
\global\advance\pmidy by -1000
\global\advance\pmidx by -1200
\put(\pmidx,\pmidy){\small $k-q_1$}
\drawline\photon[\SW\FLIPPED](\fermionbackx,\fermionbacky)[6]
\global\advance\pmidy by -200
\global\advance\photonbackx by -200
\put(\photonbackx,\pmidy){$q_1^\alpha$}
\drawline\fermion[\N\REG](\fermionbackx,\fermionbacky)[4000]
\drawarrow[\LDIR\ATTIP](\pmidx,\pmidy)
\global\advance\pmidx by -800
\put(\pmidx,\pmidy){\small $k$}
\end{picture}
\end{center}
\caption{One contribution to the 4-point photon function. The others correspond
to the six permutations of $(q_1,q_2,q_3)$.}
\label{fbox}
\end{figure}
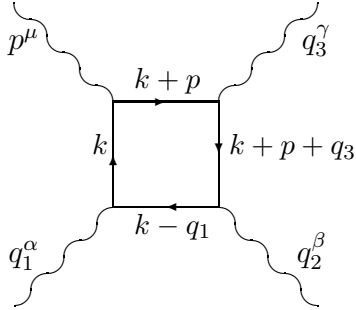

Before turning to the general proof that each term in the measure is
identically
the longitudinal projection of the related one loop graph, we will explicitly
demonstrate it for the box graph, which contains all of the essential features.
We write the four point photon graph as:
\begin{equation}
-iB^{\mu\alpha\beta\gamma}=-i(B^{\mu\alpha\beta\gamma}_1+perms.),
\end{equation}
where:
\begin{equation}
-iB^{\mu\alpha\beta\gamma}_1=
-\int\frac{dk}{(2\pi)^4}\sum_{EB}Tr[S(k+p)\gamma^\mu S(k)\gamma^\alpha
S(k-q_1)\gamma^\beta S(k+p+q_3)\gamma^\gamma],
\end{equation}
is the contribution corresponding to Fig. \ref{fbox}.
In this case, there are six permutations of final legs.
Dotting $p_\mu$ into this and reducing the traces as in the previous
Section, we get:
\begin{eqnarray}
-ip_\mu B^{\mu\alpha\beta\gamma}_1&=&
-\int\frac{dk}{(2\pi)^4}\sum_{EB}
Tr[S(p+k)\hbox{$p\mkern-8mu/\mkern 0mu$} S(k)\gamma^\alpha S(k-q_1)\gamma^\beta
S(k+p+q_3)\gamma^\gamma]\nonumber \\
&=&-\int\frac{dk}{(2\pi)^4}\sum_{EB}
\{Tr[S(k)\gamma^\alpha S(k-q_1)\gamma^\beta
S(k+p+q_3)\gamma^\gamma]\nonumber \\
&-&Tr[S(p+k)\gamma^\alpha S(k-q_1)\gamma^\beta
S(k+p+q_3)\gamma^\gamma]\}\nonumber \\
&=&-\int\frac{dk}{(2\pi)^4}
\{\sum_{ET}Tr[S(k)\gamma^\alpha S(k-q_1)\gamma^\beta
S(k+p+q_3)\gamma^\gamma]\nonumber \\
&-&\sum_{ET}Tr[S(p+k)\gamma^\alpha S(k-q_1)\gamma^\beta
S(k+p+q_3)\gamma^\gamma]\nonumber \\
&+&E^2_m(p+k)Tr[\bar{S}(k)\gamma^\alpha\bar{S}(k-q_1)\gamma^\beta
\bar{S}(k+p+q_3)\gamma^\gamma]\nonumber \\
&-&E^2_m(k)Tr[\bar{S}(p+k)\gamma^\alpha\bar{S}(k-q_1)\gamma^\beta
\bar{S}(k+p+q_3)\gamma^\gamma]\},
\end{eqnarray}
where we have separated the sum over smearing regions into two terms as before.

One can now show that the first two terms will cancel other permuted terms
via Furry's theorem.
The remaining piece is then:
\begin{eqnarray}
-ip_\mu B^{\mu\alpha\beta\gamma}&=&
-\int\frac{dk}{(2\pi)^4}
\{E^2_m(p+k)Tr[\bar{S}(k)\gamma^\alpha\bar{S}(k-q_1)\gamma^\beta
\bar{S}(k+p+q_3)\gamma^\gamma]\nonumber \\
&-&E^2_m(k)Tr[\bar{S}(p+k)\gamma^\alpha\bar{S}(k-q_1)\gamma^\beta
\bar{S}(k+p+q_3)\gamma^\gamma]\}+perms.\nonumber \\
&\stackrel{df}{=}&-ip_\mu B^{\mu\alpha\beta\gamma}_L.
\end{eqnarray}
There is also a contribution from the measure that will resurrect the identity,
coming from the transformations:
\begin{eqnarray}
\delta\psi&=&-iE^2_m\theta\bar{S}\hbox{$A\mkern-10mu/\mkern
1mu$}\bar{S}\hbox{$A\mkern-10mu/\mkern 1mu$}\bar{S}\hbox{$A\mkern-10mu/\mkern
1mu$}\psi,\quad
\delta\bar{\psi}=\bar{\psi}\hbox{$A\mkern-10mu/\mkern
1mu$}\bar{S}\hbox{$A\mkern-
10mu/\mkern 1mu$}\bar{S}\hbox{$A\mkern-10mu/\mkern 1mu$}\bar{S}\theta
E^2_mi,
\end{eqnarray}
which produces:
\begin{eqnarray}
\delta S_{meas}^{(4)}&=&-\int\frac{dp\;dq_1\;dq_2\;dq_3}{(2\pi)^8}
(2\pi)^4\delta^4(p+q_1+q_2+q_3)\int\frac{dk}{(2\pi)^4}\theta(p)\nonumber\\
&\times&\{E^2_m(k+p)Tr[\bar{S}(k)\hbox{$A\mkern-10mu/\mkern 1mu$}(q_1)
\bar{S}(k-q_1)\hbox{$A\mkern-10mu/\mkern
1mu$}(q_2)\bar{S}(k+p+q_3)\hbox{$A\mkern-10mu/\mkern 1mu$}(q_3)]\nonumber \\
&-&E^2_m(k)Tr[\hbox{$A\mkern-10mu/\mkern 1mu$}(q_3)
\bar{S}(k-q_3)\hbox{$A\mkern-10mu/\mkern
1mu$}(q_2)\bar{S}(k-q_3-q_2)\hbox{$A\mkern-10mu/\mkern
1mu$}(q_1)\bar{S}(k+p)]\}.
\end{eqnarray}
We can now symmetrize on the external photon fields to produce identical terms
with different momentum labellings:
\begin{eqnarray}
\delta S_{meas}^{(4)}&=&-\frac{1}{3!}\int\frac{dp\;dq_1\;dq_2\;dq_3}{(2\pi)^8}
(2\pi)^4\delta^4(p+q_1+q_2+q_3)\int\frac{dk}{(2\pi)^4}\theta(p)
A_\alpha(q_1)A_\beta(q_2)A_\gamma(q_3)\nonumber\\
&\times&\{E^2_m(k+p)Tr[\bar{S}(k)\gamma^\alpha
\bar{S}(k-q_1)\gamma^\beta\bar{S}(k+p+q_3)\gamma^\gamma]\nonumber \\
&-&E^2_m(k)Tr[\bar{S}(p+k)\gamma^\alpha
\bar{S}(k-q_1)\gamma^\beta\bar{S}(k+p+q_3)\gamma^\gamma]\}+perms.\nonumber \\
&=&\frac{1}{3!}\int\frac{dp\;dq_1\;dq_2\;dq_3}{(2\pi)^8}
(2\pi)^4\delta^4(p+q_1+q_2+q_3)\theta(p)A_\alpha(q_1)A_\beta(q_2)A_\gamma(q_3)
(-i)p_\mu B^{\mu\alpha\beta\gamma}_L,
\end{eqnarray}
which then gives
\begin{eqnarray}
S_{meas}^{(4)}&=&\frac{1}{4!}\int\frac{dp\;dq_1\;dq_2\;dq_3}{(2\pi)^8}
(2\pi)^4\delta^4(p+q_1+q_2+q_3)\nonumber \\
&&A_\mu(p)A_\alpha(q_1)A_\beta(q_2)A_\gamma(q_3)
B^{\mu\alpha\beta\gamma}_L,
\end{eqnarray}
which then leads to the fully corrected gauge invariant photon 4-point
function:
\begin{equation}\label{cbx}
-iB_T^{\mu\alpha\beta\gamma}=-iB^{\mu\alpha\beta\gamma}
+iB_L^{\mu\alpha\beta\gamma}.
\end{equation}
An identical construction is performed on the one loop contribution to a
general
photon n-point function in the Appendix.

\section{Higher Loop Considerations}
\label{twoloop}

We will begin by showing that the two loop correction to vacuum polarization is
gauge invariant as a consequence of the existence of the box graph contribution
to the one loop measure.
The measure corrected (transverse) box graph amplitude will be written as
(\ref{cbx}), then the three contributions to the two loop vacuum polarization
correction can now be written as:
\begin{equation}
-i\Pi_2^{\mu\gamma}(p)=
\frac{1}{2}\int\frac{dq_1}{(2\pi)^4}(2\pi)^4\delta^4(q_1+q_2)
ie\hat{D}_{\alpha\beta}(q_1)(-i)B_T^{\mu\alpha\beta\gamma}
(p,q_1,q_2,-p-q_1-q_2),
\end{equation}
since we only have `hatted' photon propagators.
The transversality of the resulting two point function then trivially follows
from the transversality of the full one loop four photon process.

It is not hard to see that this result persists for {\em all} photon n-point
functions, namely that transversality follows from the transversality of the
related single loop graph.
This holds for internal loops as well, so that any number of fermion loops may
contribute to a process, but all longitudinal photons decouple from each
separately.
Similarly, when one considers an on shell througoing fermion line, one can
rewrite the process in an analogous manner (tree graph $\times$ delta functions
and photon propagators), and infer decoupling from decoupling of the related
tree
graph.
Thus in proving the existence of the measure, we have shown that the
regularization is consistent with gauge invariance, as was to be expected.

Although this result is fairly simple in this case, it is not so easy to
implement in general.
Even if one considers the `symmetric' regularization discussed in Section
\ref{vert}, one can not so easily break up a multiloop graph into
lower loop pieces, since the parameter integral regions remain entangled in
general.
We feel that this only a technical problem and should not be impossible
to resolve.

\section{Anomalous Theories}
\label{anom}

There are two possible ways in which the construction here could fail.
The first is that the longitudinally projected vertex function has symmetries
that are destroyed when inserted into the exponentiated measure.
This is the case in the $(1+1)$-chiral Schwinger model studied by Hand
\cite{BHand}.
Briefly, the local Lagrangian (\ref{loclag}) differs by the insertion of an
axial
component in the coupling: $-\bar{\psi}\hbox{$A\mkern-10mu/\mkern 1mu$}
P_L\psi$,
where
$P_L=\frac{1}{2}(1-\gamma^5)$.
This changes the second order term in the nonlocal gauge transformation
(\ref{dlpsi}) to:
\begin{equation}
\delta\psi=-iE^2\theta\bar{S}{\hbox{$A\mkern-10mu/\mkern 1mu$}}P_L\psi,
\end{equation}
and the measure consistency condition becomes:
\begin{eqnarray}\label{Pconsis}
\delta S_{meas}^{(2)}&=&-\int\frac{dp\;dq}{(2\pi)^4}(2\pi)^4\delta^4(p+q)
\theta(p)A_\nu(q)\nonumber \\
&&\int\frac{dk}{(2\pi)^4}\{E^2(p+k)Tr[\bar{S}(k)\gamma^\nu P_L]-
E^2(k)Tr[\bar{S}(p+k)\gamma^\nu P_L]\}\nonumber \\
&=&\int\frac{dp\;dq}{(2\pi)^4}(2\pi)^4\delta^4(p+q)
\theta(p)A_\nu(q)(-i)p_\mu\Pi^{\mu\nu}_L(p).
\end{eqnarray}
We have calculated the vacuum polarization result using the same conventions as
in Fig. \ref{fvp} and identified the longitudinal
projection as in section (\ref{vp}).

This result however, is {\it not} consistent with writing:
\begin{equation}
S_{meas}^{(2)}=\frac{1}{2}\int\frac{dp\;dq}{(2\pi)^4}(2\pi)^4\delta^4(p+q)
A_\mu(p)A_\nu(q)\Pi^{\mu\nu}_L(p),
\end{equation}
since a short calculation leads to:
\begin{equation}
\Pi^{\mu\nu}_L=-\frac{4}{(2\pi)^2}\int_0^1\frac{dx}{(1+x)^2}
exp[\frac{x}{1+x}\frac{p^2}{\Lambda^2}](g^{\mu\nu}-\varepsilon^{\mu\nu}),
\end{equation}
(the antisymmetric piece arising from the two dimensional trace over
$\gamma^\mu\gamma^\nu\gamma^5$).
Clearly the action is symmetric under exchange of the field variables and the
$\varepsilon$ term is eliminated, making it impossible to satisfy the
consistency
condition (\ref{Pconsis}).

This is the case as well in the $AAA$ sector of the $U(1)$ chiral invariant
model
studied in \cite{me}.
The $VVA$ sector however, demonstrates the second possible inconsistency.
The ward identities on the triagle graph lead to two conditions:
\begin{equation}\label{axcond}
-ip_\mu\Gamma^{\mu\alpha\beta}=
-iq_{1\mu}\Gamma^{\mu\alpha\beta}=0,
\end{equation}
the first showing decoupling of the longitudinal axial vector boson, and the
second the vector boson.
The problem is that the longitudinally projected vertex functions generated
from
each of these conditions, and that appear in the measure consistency
conditions,
are not identical, and are incompatible.

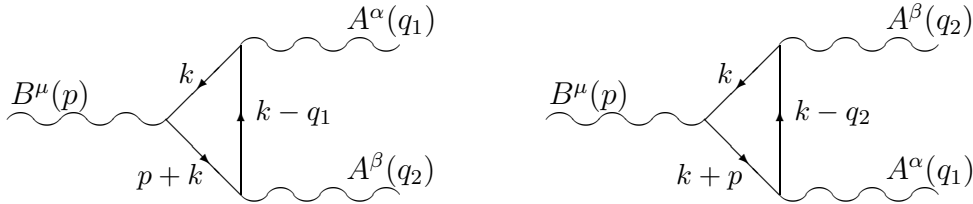
\begin{figure}[p]
\begin{center}
\begin{picture}(20000,10000)
\drawline\photon[\E\REG](1000,5000)[6]
\global\advance\photonfrontx by 100
\global\advance\photonfronty by 600
\put(\photonfrontx,\photonfronty){$B^{\mu}(p)$}
\drawline\fermion[\NE\REG](\photonbackx,\photonbacky)[4000]
\drawarrow[\SW\ATBASE](\pmidx,\pmidy)
\global\advance\pmidx by -950
\put(\pmidx,\pmidy){\small$k$}
\drawline\fermion[\SE\REG](\photonbackx,\photonbacky)[4000]
\drawarrow[\SE\ATBASE](\pmidx,\pmidy)
\global\advance\pmidx by -2500
\global\advance\pmidy by -1000
\put(\pmidx,\pmidy){\small$p+k$}
\drawline\fermion[\N\REG](\fermionbackx,\fermionbacky)[5656]
\drawarrow[\N\ATBASE](\pmidx,\pmidy)
\global\advance\pmidx by 550
\put(\pmidx,\pmidy){\small$k-q_{1}$}
\drawline\photon[\E\REG](\fermionbackx,\fermionbacky)[6]
\global\advance\pbackx by -2000
\global\advance\pbacky by 600
\put(\pbackx,\pbacky){$A^{\alpha}(q_1)$}
\drawline\photon[\E\FLIPPED](\fermionfrontx,\fermionfronty)[6]
\global\advance\pbackx by -2000
\global\advance\pbacky by 600
\put(\pbackx,\pbacky){$A^{\beta}(q_2)$}
\end{picture}
\begin{picture}(20000,10000)
\drawline\photon[\E\REG](1000,5000)[6]
\global\advance\pfrontx by 100
\global\advance\pfronty by 600
\put(\pfrontx,\pfronty){$B^{\mu}(p)$}
\drawline\fermion[\NE\REG](\photonbackx,\photonbacky)[4000]
\drawarrow[\SW\ATBASE](\pmidx,\pmidy)
\global\advance\pmidx by -950
\put(\pmidx,\pmidy){\small$k$}
\drawline\fermion[\SE\REG](\photonbackx,\photonbacky)[4000]
\drawarrow[\SE\ATBASE](\pmidx,\pmidy)
\global\advance\pmidx by -2500
\global\advance\pmidy by -1000
\put(\pmidx,\pmidy){\small$k+p$}
\drawline\fermion[\N\REG](\fermionbackx,\fermionbacky)[5656]
\drawarrow[\N\ATBASE](\pmidx,\pmidy)
\global\advance\pmidx by 550
\put(\pmidx,\pmidy){\small$k-q_2$}
\drawline\photon[\E\REG](\fermionbackx,\fermionbacky)[6]
\global\advance\pbackx by -2000
\global\advance\pbacky by 600
\put(\pbackx,\pbacky){$A^{\beta}(q_2)$}
\drawline\photon[\E\FLIPPED](\fermionfrontx,\fermionfronty)[6]
\global\advance\pbackx by -2000
\global\advance\pbacky by 600
\put(\pbackx,\pbacky){$A^{\alpha}(q_1)$}
\end{picture}
\end{center}
\caption{The two contributions to the one-loop $VVA$ sector. Note that the
axial
vector field is denoted $B^\mu$, with coupling $-i\gamma^\mu\gamma^5$}
\label{fVVA}
\end{figure}

{}From the two contributions to the $VVA$ sector in Fig. \ref{fVVA}, we find
the
longitudinal projection coming from the axial vector sector:
\begin{eqnarray}
-ip_\mu\Gamma^{\mu\alpha\beta}&=&-\int\frac{dk}{(2\pi)^4}\nonumber \\
&\times&\{E^2_m(p+k)Tr[\gamma^5\bar{S}(k)\gamma^\alpha\bar{S}(k-q_1)
\gamma^\beta]\nonumber \\
&+&E^2_m(p+k)Tr[\gamma^5\bar{S}(k)\gamma^\beta\bar{S}(k-q_2)
\gamma^\alpha]\nonumber \\
&-&E^2_m(k)Tr[\gamma^5\bar{S}(k+p)\gamma^\alpha\bar{S}(k-q_1)
\gamma^\beta]\nonumber \\
&-&E^2_m(k)Tr[\gamma^5\bar{S}(k+p)\gamma^\beta\bar{S}(k-q_2)
\gamma^\alpha]\}\nonumber \\
&=&-\frac{8p_\mu}{(4\pi)^2}
[\epsilon^{\mu\alpha\nu\beta}q_{1\nu}M(q_1;p,q_2)
+\epsilon^{\mu\beta\nu\alpha}q_{2\nu}M(q_2;p,q_1)],
\end{eqnarray}
where
\begin{equation}
M(p;q_1,q_2)=\int\frac{dx\;dy}{(1+x+y)^3}
exp[\frac{xy}{1+x+y}\frac{p^2}{\Lambda^2}
+\frac{x}{1+x+y}\frac{q_1^2}{\Lambda^2}
+\frac{y}{1+x+y}\frac{q_2^2}{\Lambda^2}],
\end{equation}
whereas that from the vector projection is:
\begin{eqnarray}
-iq_{1\alpha}\Gamma^{\mu\alpha\beta}&=&-\int\frac{dk}{(2\pi)^4}\nonumber \\
&&\{E^2_m(k)Tr[\gamma^5\bar{S}(k-q_1)\gamma^\beta\bar{S}(k+p)
\gamma^\mu]\nonumber \\
&+&E^2_m(k-q_1)Tr[\gamma^5\bar{S}(k+p)\gamma^\beta\bar{S}(k)
\gamma^\mu]\nonumber
\\
&-&E^2_m(k-q_1)Tr[\gamma^5\bar{S}(k)\gamma^\beta\bar{S}(k+p)
\gamma^\mu]\nonumber
\\
&-&E^2_m(k)Tr[\gamma^5\bar{S}(k+p)\gamma^\beta\bar{S}(k-q_1)
\gamma^\mu]\nonumber
\\
&=&-\frac{8q_{1\alpha}}{(4\pi)^2}
[\epsilon^{\mu\alpha\nu\beta}q_{2\nu}M(q_2;p,q_1)
+\epsilon^{\mu\beta\nu\alpha}p_\nu M(p;q_1,q_2)].
\end{eqnarray}
It is not hard to see from this that it is impossible to write a measure
correction that will satisfy (\ref{axcond}).

This shows that one need not attempt to calculate the measure directly from the
nonlocal gauge transformations in order to determine wether a theory is
anomalous
or not.
It is sufficient (and equivalent) to check the Ward-Takahashi identities on
graphs where possible conflicts of this type may arise.

\section*{Conclusions}

We have shown that nonlocal regulated QED has a one loop invariant measure to
all
orders through directly equating it to the related longitudinally projected
vertex functions.
This result then leads to decoupling of longitudinal photons from all processes
with on-shell external fermions, and so we have proven that we will generate a
gauge invariant perturbation series.

Clearly the same considerations will hold in other theories as well, and one
may
state with some confidence that if one can consistently introduce a measure
into
the generating functional solely on the basis of imposing the Ward-Takahashi
identities, then this is indeed the required invariant measure.

\section*{Acknowledgements}

The author wishes to thank L. Demopoulos and J. W. Moffat for stimulating
discussions, and the Natural Sciences and Engineering Research Council Canada
for
financial support.

\appendix
\section{One Loop QED Measure}

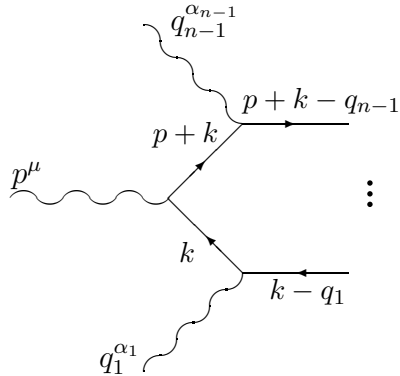
\begin{figure}[p]
\begin{center}
\begin{picture}(10000,15000)
\drawline\photon[\SE\REG](3000,14000)[6]
\global\advance\photonfrontx by 1000
\global\advance\photonfronty by -200
\put(\photonfrontx,\photonfronty){$q_{n-1}^{\alpha_{n-1}}$}
\drawline\fermion[\E\REG](\photonbackx,\photonbacky)[4000]
\drawarrow[\E\ATTIP](\pmidx,\pmidy)
\global\advance\pmidx by -2000
\global\advance\pmidy by 600
\put(\pmidx,\pmidy){\small $p+k-q_{n-1}$}
\drawline\fermion[\SW\REG](\photonbackx,\photonbacky)[4000]
\drawarrow[\NE\ATTIP](\pmidx,\pmidy)
\global\advance\pmidx by -2000
\global\advance\pmidy by 700
\put(\pmidx,\pmidy){\small$p+k$}
\drawline\photon[\W\REG](\fermionbackx,\fermionbacky)[6]
\global\advance\photonbackx by 100
\global\advance\photonbacky by 600
\put(\photonbackx,\photonbacky){$p^\mu$}
\drawline\fermion[\SE\REG](\fermionbackx,\fermionbacky)[4000]
\drawarrow[\NW\ATTIP](\pmidx,\pmidy)
\global\advance\pmidx by -1000
\global\advance\pmidy by -1000
\put(\pmidx,\pmidy){\small$k$}
\drawline\photon[\SW\REG](\fermionbackx,\fermionbacky)[6]
\global\advance\pbackx by -1700
\global\advance\pbacky by 200
\put(\pbackx,\pbacky){$q_1^{\alpha_1}$}
\drawline\fermion[\E\REG](\fermionbackx,\fermionbacky)[4000]
\drawarrow[\W\ATTIP](\pmidx,\pmidy)
\global\advance\pmidx by -1000
\global\advance\pmidy by -1000
\put(\pmidx,\pmidy){\small $k-q_1$}
\global\advance\fermionbacky by 2500
\global\advance\fermionbackx by 500
\put(\fermionbackx,\fermionbacky){\huge $\vdots$}
\end{picture}
\end{center}
\caption{A section of one contribution to the general n-point photon graph.}
\label{fnpt}
\end{figure}

Consider first the n-point photon graph, a portion of which is shown in Fig.
\ref{fnpt}:
\begin{equation}
-iN^{\mu\alpha_1\alpha_2\ldots\alpha_{n-1}}
=-i\sum_{P}N^{\mu\alpha_1\alpha_2\ldots\alpha_{n-1}}_{P},
\end{equation}
where the sum is over permutations of $\alpha_i$.
(Note that we could just as easily permute say $\mu\alpha_2\ldots\alpha_{n-1}$
and recover the same result. This means that
$p_\mu N^{\mu\alpha_1\ldots\alpha_{n-1}}=q_{\alpha_1}
N^{\mu\alpha_1\alpha_2\ldots\alpha_{n-1}}$, a result that will be important
later.)

We find for the identity permutation:
\begin{eqnarray}
-ip_\mu N_1^{\mu\alpha_1\ldots\alpha_{n-1}}
&=&-\int\frac{dk}{(2\pi)^4}\sum_{En}\{Tr[S(k)\gamma^{\alpha_1}S(k-q_1)
\gamma^{\alpha_2}\cdots
\gamma^{\alpha_{n-2}}S(k+p+q_{n-1})\gamma^{\alpha_{n-1}}]\nonumber \\
&-&Tr[S(k+p)\gamma^{\alpha_1}S(k-q_1)\gamma^{\alpha_2}\cdots
\gamma^{\alpha_{n-2}}S(k+p+q_{n-1})\gamma^{\alpha_{n-1}}].
\end{eqnarray}
First we note that the odd order graphs will disappear by Furry's theorem
(one can see this by considering the permutation that replaces $q_m$ by
$q_{n-m}$, and noting that the two terms differ by the reversal of the fermion
line, and hence must cancel).
Then we expand the trace (as before) into a term that gives a contribution to
an
$n-1$ order graph and a term with barred propagators:
\begin{eqnarray}
-ip_\mu N_1^{\mu\alpha_1\ldots\alpha_{n-1}}&=&-\int\frac{dk}{(2\pi)^4}\nonumber
\\
&&\{\sum_{E(n-1)}Tr[S(k)\gamma^{\alpha_1}S(k-q_1)\gamma^{\alpha_2}\cdots
\gamma^{\alpha_{n-2}}S(k+p+q_{n-1})\gamma^{\alpha_{n-1}}]\nonumber \\
&-&\sum_{E(n-1)}Tr[S(k+p)\gamma^{\alpha_1}S(k-q_1)\gamma^{\alpha_2}\cdots
\gamma^{\alpha_{n-2}}S(k+p+q_{n-1})\gamma^{\alpha_{n-1}}]\nonumber \\
&+&E^2_m(k+p)Tr[\bar{S}(k)\gamma^{\alpha_1}\bar{S}(k-q_1)\gamma^{\alpha_2}\cdots
\gamma^{\alpha_{n-2}}\bar{S}(k+p+q_{n-1})\gamma^{\alpha_{n-1}}]\nonumber \\
&-&E^2_m(k)Tr[\bar{S}(k+p)\gamma^{\alpha_1}\bar{S}(k-q_1)\gamma^{\alpha_2}\cdots
\gamma^{\alpha_{n-2}}\bar{S}(k+p+q_{n-1})\gamma^{\alpha_{n-1}}].
\end{eqnarray}
The first two terms are contributions to $n-1$ order graphs and will disappear
when symmetrized, since $n-1$ is now odd.
We are then left with;
\begin{eqnarray}
-ip_\mu N_1^{\mu\alpha_1\ldots\alpha_{n-1}}&=&-\int\frac{dk}{(2\pi)^4}\nonumber
\\
&+&E^2_m(k+p)Tr[\bar{S}(k)\gamma^{\alpha_1}\bar{S}(k-q_1)\gamma^{\alpha_2}\cdots
\gamma^{\alpha_{n-2}}\bar{S}(k+p+q_{n-1})\gamma^{\alpha_{n-1}}]\nonumber \\
&-&E^2_m(k)Tr[\bar{S}(k+p)\gamma^{\alpha_1}\bar{S}(k-q_1)\gamma^{\alpha_2}\cdots
\gamma^{\alpha_{n-2}}\bar{S}(k+p+q_{n-1})\gamma^{\alpha_{n-1}}],
\end{eqnarray}
and so we define:
\begin{equation}
-ip_\mu N^{\mu\alpha_1\ldots\alpha_{n-1}}_L\stackrel{df}{=}-ip_\mu\sum_P
N_{P}^{\mu\alpha_1\ldots\alpha_{n-1}}.
\end{equation}

This is related to the calculation of the measure as follows.
Consider the gauge transformations containing $n-1$ photon fields:
\begin{equation}
\delta\psi=-iE^2_m\theta(\bar{S}\hbox{$A\mkern-10mu/\mkern 1mu$})^{n-1}\psi.
\end{equation}
Calculating the measure consistency condition as in (\ref{mcon}), we find that
this gives:
\begin{eqnarray}
\delta S^{(n)}_{meas}&=&-\int\frac{dp\;dq_1\ldots dq_{n-1}}{(2\pi)^{2n}}
(2\pi)^4\delta^4(p+q_1+\ldots+q_{n-1})\nonumber \\
&&\int\frac{dk}{(2\pi)^4}\theta(p)E^2_m(k+p)\nonumber \\
&&Tr[\bar{S}(k)\hbox{$A\mkern-10mu/\mkern
1mu$}(q_1)\bar{S}(k-q_1)\hbox{$A\mkern-
10mu/\mkern 1mu$}(q_2)\ldots
\hbox{$A\mkern-10mu/\mkern 1mu$}(q_{n-2})\bar{S}(k+p+q_{n-1})\hbox{$A\mkern-
10mu/\mkern 1mu$}(q_{n-1})]\nonumber \\
&=&-\frac{1}{(n-1)!}\int\frac{dp\;dq_1\ldots dq_{n-1}}{(2\pi)^{2n}}
(2\pi)^4\delta^4(p+q_1+\ldots+q_{n-1})\nonumber \\
&&A_{\alpha_1}(q_1)A_{\alpha_2}(q_2)\ldots A_{\alpha_{n-1}}(q_{n-1})
\sum_{P}\int\frac{dk}{(2\pi)^4}\theta(p)E^2_m(k+p)\nonumber \\
&&Tr[\bar{S}(k)\gamma^{\alpha_1}\bar{S}(k-q_1)\gamma^{\alpha_2}\ldots
\gamma^{\alpha_{n-2}}\bar{S}(k+p+q_{n-1})\gamma^{\alpha_{n-1}}],
\end{eqnarray}
and once the conjugate is added to this we find the full condition on the
measure:
\begin{eqnarray}
\delta S^{(n)}_{meas}&=&-\frac{1}{(n-1)!}\int
\frac{dp\;dq_1\ldots dq_{n-1}}{(2\pi)^{2n}}
(2\pi)^4\delta^4(p+q_1+\ldots+q_{n-1})\nonumber \\
&&A_{\alpha_1}(q_1)A_{\alpha_2}(q_2)\ldots A_{\alpha_{n-1}}(q_{n-1})
\sum_{P}\int\frac{dk}{(2\pi)^4}\theta(p)\nonumber \\
&&\{E^2_m(k+p)Tr[\bar{S}(k)\gamma^{\alpha_1}\bar{S}(k-q_1)\gamma^{\alpha_2}
\ldots\gamma^{\alpha_{n-2}}\bar{S}(k+p+q_{n-1})\gamma^{\alpha_{n-1}}]
\nonumber \\
&-&E^2_m(k)Tr[\bar{S}(k+p)\gamma^{\alpha_1}\bar{S}(k-q_1)\gamma^{\alpha_2}\ldots
\gamma^{\alpha_{n-2}}\bar{S}(k+p+q_{n-1})\gamma^{\alpha_{n-1}}]\}
\nonumber \\
&=&\frac{1}{(n-1)!}\int\frac{dp\;dq_1\ldots dq_{n-1}}{(2\pi)^{2n}}
(2\pi)^4\delta^4(p+q_1+\ldots+q_{n-1})\nonumber \\
&&\theta(p)
A_{\alpha_1}(q_1)A_{\alpha_2}(q_2)\ldots A_{\alpha_{n-1}}(q_{n-1})
(-i)p_\mu N^{\mu\alpha_1\ldots\alpha_{n-1}}_L.
\end{eqnarray}
Due to the above stated symmetry, we can then immediately write:
\begin{eqnarray}
S^{(n)}_{meas}&=&\frac{1}{n!}\int\frac{dp\;dq_1\ldots dq_{n-1}}{(2\pi)^{2n}}
(2\pi)^4\delta^4(p+q_1+\ldots+q_{n-1})\nonumber \\
&&\int\frac{dk}{(2\pi)^4}A_\mu(p)
A_{\alpha_1}(q_1)A_{\alpha_2}(q_2)\ldots A_{\alpha_{n-1}}(q_{n-1})
N^{\mu\alpha_1\ldots\alpha_{n-1}}_L.
\end{eqnarray}
The resulting full n-point function:
\begin{equation}
-iN_T^{\mu\alpha_1\ldots\alpha_{n-1}}=
-iN^{\mu\alpha_1\ldots\alpha_{n-1}}+iN_L^{\mu\alpha_1\ldots\alpha_{n-1}},
\end{equation}
is then transverse.

At order n, the local graph diverges as $(p^2)^{2-n/2}$ and so the measure
contribution at the same order here will be proportional to
$(\Lambda^2)^{2-n/2}$.
This indicates that nothing beyond $n=4$ will survive in the local limit, which
is in accord with the fact that counterterms are not necessary beyond fourth
order in local regularization schemes.
Note that this result relates the measure directly to the one loop graph that
it
is required to `fix up', hence only even orders appear in the measure.
One may wonder about wether the resulting amplitude is an entire function of
the
4-momentum invariants of the process in question, but since the term explicitly
comes from a convergent integral over barred propagators (that have no pole),
we
will not encounter any singularities when passing to Minkowski spacetime.
We have also not had to resort to putting any external fields on shell, so that
longitudinal photons will also decouple from internal fermion loops as well.

%
%
%
%
%
%
%
%
%
%
\end{document}